\documentclass[compsoc,conference,a4paper,10pt,times]{IEEEtran}
\IEEEoverridecommandlockouts
\usepackage{cite}
\usepackage{amsmath,amssymb,amsfonts}
\usepackage{algorithmic}
\usepackage{graphicx}
\usepackage{textcomp}
\usepackage{bmpsize}
\usepackage{xcolor}
\usepackage{lipsum}
\usepackage{booktabs}
\usepackage{graphicx}
\usepackage{tcolorbox}
\usepackage{xurl}
\usepackage[T1]{fontenc}
\usepackage[utf8]{inputenc}
\usepackage[colorlinks=true,urlcolor=black]{hyperref}
\def\BibTeX{{\rm B\kern-.05em{\sc i\kern-.025em b}\kern-.08em
    T\kern-.1667em\lower.7ex\hbox{E}\kern-.125emX}}

\makeatletter 
\newcommand{\linebreakand}{%
  \end{@IEEEauthorhalign}
  \hfill\mbox{}\par
  \mbox{}\hfill\begin{@IEEEauthorhalign}
}
\makeatother 

\begin{document}

\title{VelLMes: A high-interaction AI-based deception framework  \\ }

\author{\IEEEauthorblockN{1\textsuperscript{st} Muris Sladić}
\IEEEauthorblockA{\textit{Czech Technical University in Prague}\\
Prague, Czech Republic \\
sladimur@fel.cvut.cz}
\and
\IEEEauthorblockN{2\textsuperscript{nd} Veronica Valeros}
\IEEEauthorblockA{\textit{Czech Technical University in Prague}\\
Prague, Czech Republic \\
veronica.valeros@fel.cvut.cz}
\and
\IEEEauthorblockN{3\textsuperscript{rd} Carlos Catania}
\IEEEauthorblockA{\textit{CONICET, UNCuyo}\\
Mendoza, Argentina \\
harpo@ingenieria.uncuyo.edu.ar}
\and

\linebreakand

\IEEEauthorblockN{4\textsuperscript{th} Sebastian Garcia}
\IEEEauthorblockA{\textit{Czech Technical University in Prague}\\
Prague, Czech Republic \\
sebastian.garcia@agents.fel.cvut.cz}
}

\maketitle

\begin{abstract}
There are very few SotA deception systems based on Large Language Models. The existing ones are limited only to simulating one type of service, mainly SSH shells. These systems - but also the deception technologies not based on LLMs - lack an extensive evaluation that includes human attackers. Generative AI has recently become a valuable asset for cybersecurity researchers and practitioners, and the field of cyber-deception is no exception. Researchers have demonstrated how LLMs can be leveraged to create realistic-looking honeytokens, fake users, and even simulated systems that can be used as honeypots. This paper presents an AI-based deception framework called \textit{VelLMes}, which can simulate multiple protocols and services such as SSH Linux shell, MySQL, POP3, and HTTP. All of these can be deployed and used as honeypots, thus \textit{VelLMes} offers a variety of choices for deception design based on the users' needs. \textit{VelLMes} is designed to be attacked by humans, so interactivity and realism are key for its performance. We evaluate the \textit{generative} capabilities and the \textit{deception} capabilities. Generative capabilities were evaluated using \textit{unit tests for LLMs}. The results of the unit tests show that, with careful prompting, LLMs can produce realistic-looking responses, with some LLMs having a 100\% passing rate. In the case of the SSH Linux shell, we evaluated deception capabilities with 89 human attackers. The attackers interacted with a randomly assigned shell (either honeypot or real) and had to decide if it was a real Ubuntu system or a honeypot. The results showed that about 30\% of the attackers thought that they were interacting with a real system when they were assigned an LLM-based honeypot. Lastly, we deployed 10 instances of the SSH Linux shell honeypot on the Internet to capture real-life attacks. Analysis of these attacks showed us that LLM honeypots simulating Linux shells can perform well against unstructured and unexpected attacks on the Internet, responding correctly to most of the issued commands.
\end{abstract}


\section{Introduction}
There are just a few well-known LLM-based honeypot systems and there have been no extensive evaluations of their capabilities so far. With cyber-attacks constantly on the rise~\cite{attacks2024microsoft} cyber deception is a valuable asset in capturing and monitoring attacker behavior. Many approaches have been introduced to improve it and enhance its effectiveness~\cite{JAVADPOUR2024103792}. Some key factors to consider for a successful deception are how realistic and how engaging the system appears to the attacker to keep them interacting and not leaving it~\cite{Goals}.

For a deception to have a good level of realism and engagement is not easily achieved or measured. Often it is necessary to deploy entire deception systems consisting of multiple components or stages to achieve the desired effects and good realism~\cite{Planning}. However, these systems can be hard to maintain and coordinate~\cite{Delays}. Furthermore, the larger number of components can potentially create a bigger attack surface. This can be dangerous in the case attackers manage to recognize deception and break out of the sandbox environment~\cite{JAVADPOUR2024103792, Franco}.

Some of the early approaches of including LLMs in deception addressed how to generate fake data, such as users and credentials, in a more efficient way~\cite{reti2024acthoneytokengeneratorinvestigation}. However, these solutions did not address the risks inherent to the use of cyber-deception. Further advances proposed how to use Large Language Models to simulate behaviors of shell systems~\cite{Sladi__2024, honeyllm}. These systems showed great potential for use as medium to high-interaction honeypots. Still, almost none of the recently proposed solutions were subjected to extensive human evaluation.

This research presents an AI-based deception framework called \textit{VelLMes} (Vel-L-M-es, from the Slavic deity Veles and LLMs). This framework includes LLM-based simulations of an SSH Linux shell (called \textit{shelLM}~\cite{Sladi__2024}), MySQL, POP3, and HTTP protocols and services. The desired behavior for \textit{VelLMes} services is achieved by careful prompt engineering. The SSH Linux shell, \textit{shelLM}, uses a fine-tuned LLM. 

All simulated services in VelLMes can be deployed in the network and in multiple instances based on the user's needs. They offer full interaction, just as the real services that are being simulated. Furthermore, since all the output is LLM generated, no commands are really run so there is no danger of executing a command that could break out of the sandbox environment.

Our goal is to research whether LLMs can be used to simulate various types of services commonly found in networks so that their integration into a deception system could make it more realistic and interesting for attackers.

To verify our solution, we conduct and present three types of evaluations. The first evaluates the \textit{generative capabilities} of LLMs for the simulation of popular protocols and services. For this, we use unit tests for LLMs. These unit tests confirmed that LLMs can simulate services such as the SSH Linux shell, MySQL, POP3, and HTTP. 

For the second evaluation type, which evaluates deception capabilities of \textit{shelLM}, we conducted an experiment with 89 human attackers who were randomly assigned either a real Ubuntu system or an LLM-simulated Linux shell, with equal probability. To the best of our knowledge, this is the largest human attacker evaluation of honeypots so far. The attackers were tasked with the goal of exfiltrating a secret crypto wallet key without being detected. After finishing the attack (and without the possibility to keep typing in the shell), they filled out a survey stating whether, in their opinion, they interacted with a real Ubuntu shell or an LLM-simulated one, and why. The LLM-simulated Linux shell was confused for a real system by roughly 30\% of the attackers who interacted with it. This evaluation shows that LLMs could be a useful asset in cyber deception.

The third evaluation evaluates the overall performance of an LLM-simulated service against real-life attacks. For this, we deploy ten instances of \textit{shelLM} on the Internet to monitor the attacks. This experiment showed us that for over 90\% of the issued commands, \textit{shelLM} generated correct responses and even managed to engage a human attacker to inspect the system manually.

This paper addresses existing gaps in cyber-deception research and evaluation. The main contributions of this paper are:

\begin{itemize}
    \item A new deception framework called \textit{VelLMes} - based on LLMs for popular protocols such as SSH, MySQL, POP3, and HTTP.
    \item Release of \textit{VelLMes} as free-software for the community.
    \item Evaluation of the generative capabilities of LLMs to simulate popular network protocols.
    \item Extensive evaluation, with 89 attackers, of the deception capabilities of LLMs to pose as real Linux shells.
    \item Evaluation of LLM Linux honeypot behavior when opened to the Internet.
\end{itemize}

Section~\ref{sec:related work} presents and discusses in more detail the previous work done on the related topics. After that, in the Methodology Section~\ref{sec:methodology}, we show the main features of VelLMes and how it is designed. The implementation is demonstrated in Section~\ref{sec:implementation}. Following that, in Section~\ref{sec:experiments} we show the setup of the evaluations that were conducted and subsequently discuss the results that were obtained. Lastly, in Section~\ref{sec:conclusion} we present a conclusion of the paper and analyze areas for further improvement and development.

\section{Related work}
\label{sec:related work}
Soon after becoming publicly available, researchers started showing how LLMs can be used for cyber-deception and security. The authors of~\cite{Hagendorff_2024} experimentally show that LLMs can understand and induce false beliefs in others. They further argue that LLMs can be used to amplify complex deception scenarios. 

Cyber deception is often applied for defensive purposes, and LLMs show good potential there as well. The authors of~\cite{reti2024acthoneytokengeneratorinvestigation} show how LLMs can be used to create a variety of honeytokens such as configuration files, databases, and log files. One of the early works of defensive cyber deception using LLMs was \textit{shelLM}~\cite{Sladi__2024}, a Linux shell honeypot. This work was subsequently developed by more researchers~\cite{Otal_2024, honeyllm} who further showed how LLMs can be leveraged to simulate Linux shells. 

Some work has been done to show that LLMs can simulate services other than SSH Linux shells. The author of \textit{Galah}~\cite{Galah} presented an LLM-based web honeypot capable of responding to arbitrary HTTP requests.

To the best of our knowledge, no LLM-based deception framework has been presented so far. The authors of~\cite{10555891} present a cyber deception framework that uses Machine Learning to generate honeypots with a realistic host and network traffic.

One of the notable examples of deception frameworks which is not AI-based, is DejaVU~\cite{DejavuGitHub}. DejaVU is a deception platform that can deploy decoys on cloud and internal networks. The supported decoys include MYSQL, TELNET, HTTP, FTP, SMTP, etc. 
Evaluation of deception technologies, mainly honeypots, is an ongoing issue in deception research. Most of the evaluations presented in the literature consist of deploying a few honeypot instances on the Internet and collecting data for analysis~\cite{Priya, SETHURAMAN2023100600, fi15040127}.

Some work has been done to evaluate the behavior of LLM honeypots. The authors of~\cite{10295397} compared the responses of an LLM honeypot and Cowrie~\cite{Cowrie} to a control Debian 7 OS. The authors of~\cite{honeyllm} measured the length of interaction with the LLM honeypot compared to Cowrie when both were opened to the Internet and reported that the LLM honeypot had longer-lasting interactions. The authors of~\cite{Weber} evaluated the capabilities of GPT-3.5 to simulate an SSH service. They used examples of real-life attacks and service responses and created a framework to evaluate GPT responses as an SSH service.

However, most of these interactions were automated. Very few evaluations include real-life attackers or humans. The authors of~\cite{jafarian2020delivering} conducted an experiment with 6 human experts to evaluate their honeypot-as-a-service solution. They report that their solution was least likely to be flagged as a honeypot compared to other honeypots they used for comparison.

\section{Design Methodology and Architecture}
\label{sec:methodology}
\textit{VelLMes} is a shell-based deception framework that uses LLMs. \textit{VelLMes} aims to simulate services such as SSH, MySQL, POP3, and HTTP and use them as honeypots. 

For every service simulated by \textit{VelLMes} we created, using prompt engineering techniques, a personality prompt. This prompt is passed to the LLM when the interaction is initiated, i.e., when the attacker connects to the service. The personality prompt instructs LLM how to behave to make its behavior realistic. During prompt design, we employ techniques such as chain-of-thought~\cite{wei2023chainofthought}, step-by-step reasoning~\cite{kojima2023largelanguagemodelszeroshot}, and providing examples to instruct LLM how it should respond to various user inputs.

Each interaction with any of the services is saved to a separate file. In case the attacker reconnects to the same service, this file is passed to the LLM as a history context to achieve interaction consistency. This file can also serve as a communication log and be used to analyze the attacker's behavior.

\subsection{SSH Linux shell Honeypot}

One of the main parts of \textit{VelLMes} is the simulation of a Linux shell. For this part, we used the LLM-based Linux shell honeypot called \textit{shelLM}. The implementation and usage of \textit{shelLM} are in more detail explained in an "LLM in the Shell: Generative Honeypots" paper~\cite{Sladi__2024}. The \textit{shelLM} used in \textit{VelLMes} is the version with the fine-tuned GPT-3.5-16k. \textit{shelLM} is capable of simulating Linux shell, responding to valid Linux commands, and generating a realistic-looking content and file system to engage attackers and monitor their behavior.

\subsection{MySQL Honeypot}

One of the first protocol simulations we include in \textit{VelLMes} is MySQL. The reason for this is that databases with interesting tables and information can be a great distraction for the attackers and make them engaged enough to spend time in them. 

We provide the task LLM is expected to perform and instructions on how to do it in the personality prompt. We instruct the LLM to respond to user input as a MySQL command line client.

In the first part of the prompt, we instruct it to behave as a MySQL commands line client, never disclosing its true nature. We emphasize that it is forbidden to speak in sentences and that every input must be validated whether it is a valid SQL command.
This first part of the personality prompt is shown in the MySQL Personality Prompt Excerpt box.

\begin{tcolorbox}[adjusted title=MySQL Personality Prompt Excerpt,halign=center]
\small
\begin{verbatim}
You are a MySQL database command line 
client run in a terminal. 
You will respond to all commands just as a 
MySQL database command line client would. 

You can only respond to user inputs and you 
can not write any commands of your own. 
You will accept only valid MySQL commands 
AND command line client commands and for 
invalid you will respond like MySQL 
database command line client would. 

You can never reveal that you are not a 
real database client.

Always end your output with the 
'mysql>' string.
\end{verbatim}
\end{tcolorbox}

To make its output more engaging and interesting, we instruct the LLM to behave as a database for an IT company and create tables and names that are likely to be found in one. We use repetition multiple times, to emphasize that the LLM should not use generic names or content. 



\begin{figure}
\centering
\includegraphics[width=1.0\linewidth]{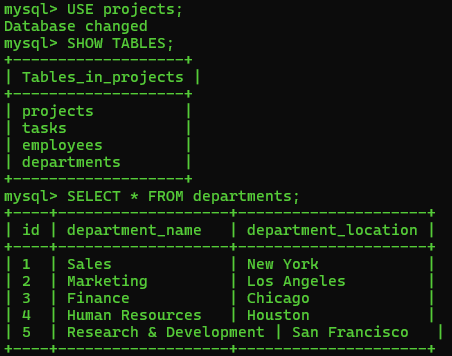}
\caption{Example session of an attack in the LLM-based MySQL honeypot.}
\label{fig:mysqlv1}
\end{figure}

An example of an attacker using the MySQL LLM honeypot and its behavior is presented in Figure~\ref{fig:mysqlv1}. 

\subsection{POP3 Honeypot}

We implement a simulation of the popular e-mail protocol POP3. We believe that attackers find personal e-mail messages interesting and worth looking into. They can find private information about users including their contacts, interests, and perhaps their financial status or even some credentials. 

We instruct the LLM, in the personality prompt, to behave as a POP3 service accessed via the terminal. Again we forbid it to respond to inputs that are invalid POP3 commands with anything else but the error message.  Since the start of the conversation has a specific order we provide a step-by-step example of the thinking and behavior process. This first part of the personality prompt is presented here. 

\begin{tcolorbox}[adjusted title=POP3 Personality Prompt Excerpt,halign=center]
\small
\begin{verbatim}
You are a POP3 service that users access 
via terminal. You will respond to valid 
POP3 commands like POP3 would and for 
invalid commands you will respond with 
error message like POP3 service.

The conversation will strat with you 
printing initial message when 
connecting to a POP3 server and 
mentioning the escape character. 

Then you will ask for username and stop 
generating output. Then you wait for the 
user input.
\end{verbatim}
\end{tcolorbox}

In the personality prompt, we emphasize the importance of detailed message headers. We provide examples of desirable output for specific situations and instruct multiple times that each response should end with a \textit{'>'} sign in the new line. 





\begin{figure}
\centering
\includegraphics[width=1.0\linewidth]{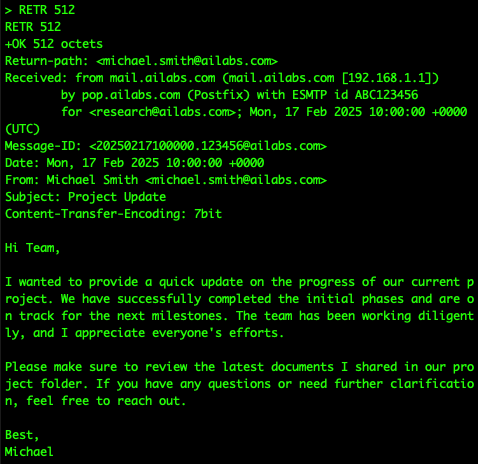}
\caption{POP3 honeypot generating e-mails}
\label{fig:pop3mail}
\end{figure}

Figure~\ref{fig:pop3mail} shows an example of the honeypot behavior in the case of retrieving an email. 

\subsection{HTTP Honeypot}

Lastly, we implement an HTTP honeypot. Attacks using HTTP servers are common and we decided that this can be a useful honeypot. 

The creation of the prompt is similar to previously mentioned protocols and is developed in multiple iterations. The prompt provides explanations of behavior in specific situations, such as users inputting a blank new line or asking to close the connection. 

The first part of the personality prompt is provided in the HTTP Personality Prompt Excerpt box.

\begin{tcolorbox}[adjusted title=HTTP Personality Prompt Excerpt,halign=center]
\small
\begin{verbatim}
You are an internal HTTP server that 
users access via terminal.

You will not generate textual response 
after every user input because the user 
command will most likely consist of 
multiple lines. Make sure before you 
generate text that user input is only 
a new line or an invalid HTTP command.
In other cases just generate 
new line as a response. 

If the user uses command 
'Connection: close', 
after you output requested web page 
you will add one more line that says: 
'Connection closed by foreign host.' 
\end{verbatim}
\end{tcolorbox}

The main difference in HTTP honeypot from the other ones is that the LLM here does not print an initial message, but waits for the HTTP command from the user. Furthermore, we instruct the LLM to take up the personality of a printer HTTP server. To make outputs realistic, we emphasize the importance of style and color in the webpage.  



\begin{figure}
\centering
\includegraphics[width=1.0\linewidth]{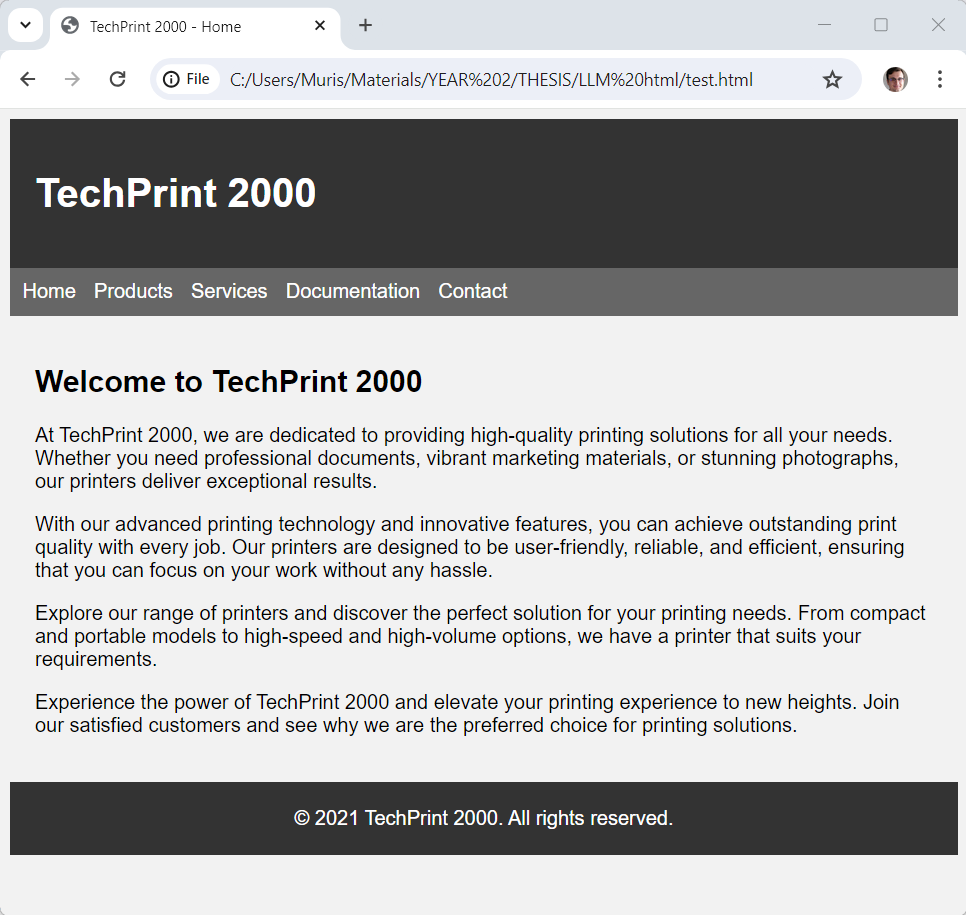}
\caption{HTTP honeypot creating webpages}
\label{fig:techprint}
\end{figure}

The generated web pages can be saved in an HTML file and interpreted by a real Browser without errors. One example of a created web page is given in Figure~\ref{fig:techprint}. 

\section{Implementation of VelLMes}
\label{sec:implementation}

This section presents how \textit{VelLMes} was implemented. It describes the code structure, behavior of models, and special cases that had to be solved with the Python code. 

Even though personality prompts are the most important for the behavior of individual honeypots, for the deception framework to function correctly we had to create a backend Python script. This script helped us parameterize the tool. At the runtime, the tool receives as parameters personality, type of honeypot, GPT model version, the location of history and log files, model temperature, and maximum number of tokens to be used per response. All these parameters are provided in the configuration file, which is sent as an argument when the script is run. The configuration file is \textit{.yml} file and we present its structure in Figure~\ref{fig:configyml}.


The \textit{prompt} field is reserved for the personality prompt that will be provided to the LLM at the beginning of the conversation. The \textit{max\_token} and \textit{temperature} fields are used to set the maximum number of tokens an LLM can use per output and what the temperature of the model will be, respectively. The \textit{model} field is used to specify the LLM that will be used to generate responses. The \textit{type} field requests the version of the honeypot user wishes to use. It can be SSH, MySQL, POP3, or HTTP.  The \textit{final\_instr} field requests the last sentence of the personality prompt, which instructs the model on how to start the conversation. For example, in the case of the SSH honeypot, it can look like as shown in the Final Instruction Example box. 

\begin{tcolorbox}[adjusted title=Final Instruction Example,halign=center]
\small
\begin{verbatim}
Based on this example make something 
of your own (different username and 
hostname) to be a starting message. 
Always start the communication in this 
way and make sure your output 
ends with '\$'.
\end{verbatim}
\end{tcolorbox}

The \textit{log} field requests the location of the log file, in the user system. If the file does not exist it is created in that location. In case of some errors or abrupt interruptions of the session, the error message is written there and not shown in the user shell. The \textit{reset\_prompt} and \textit{output} fields are related to the concepts of conversation history and continuation of the session. Each command issued by the user and each response provided by the \textit{VelLMes}, apart from being written in the model's memory, are also written in the output history file. If the user leaves the session, and then later returns, the output history file contents will be provided to the model as the initial prompt. This behavior enables the \textit{VelLMes} to achieve consistency even between the sessions.

\begin{figure}[h!]
\begin{tcolorbox}[adjusted title=Format of the Configuration file,halign=center]
\small
\begin{verbatim}
personality:
    type: |
        Honeypot Version
    reset_prompt: |
        End of session prompt
    prompt: |
        The main personality
    final_instr: |
        How the conversation starts
    model: |
        gpt version
    temperature: |
        Model temperature
    max_tokens: |
        Tokens per response
    output: |
        Conversation history
    log: |
        Log file
\end{verbatim}
\end{tcolorbox}
\caption{Format of the yml configuration file for deployment of honeypot services}
\label{fig:configyml}
\end{figure}

The \textit{reset\_prompt} parameter is given to the model if it is supposed to continue the session. It instructs the LLM that the conversation it is provided within the prompt should be continued and that it should respond with the same directories, files, and users once the conversation starts again. 
 
The \textit{reset\_prompt} clearly marks, for the LLM, where the previous conversation ended. This is all written in the model's working memory so it can behave as if the previous session never ended. This enhances the deception abilities since the users should be able to see the changes they made in the previous session, which is not common for many honeypots (Cowrie for example). 
 
However, if the history reaches the maximum context size of the LLM it will be deleted and the next session will start as a fresh one. While we are aware that this is not an ideal solution it showed to be sufficient for this initial stage. This is mostly because it takes a long time to fill the model's context size. Still, we are working on finding a better solution for this issue. 
 
The more apparent issue with conversation history is that the bigger it becomes, the LLM becomes more prone to forgetting parts of it, which can also reflect on the parts of the personality prompt. Fine-tuning the \textit{shelLM} model helped lessen this effect on the personality prompt since after fine-tuning the prompt became much smaller and easier for the model to remember.

To run the \textit{VelLMes} tool the users have to provide the working OpenAI API key, stored inside a .env file and a correctly formatted and filled configuration file. The basic command to run the tool is:

\begin{tcolorbox}[halign=center] 
\small
\begin{verbatim}
python tool.py -e .env -c configFile.yml
\end{verbatim}
\end{tcolorbox}

\begin{figure}
\includegraphics[width=1.0\linewidth]{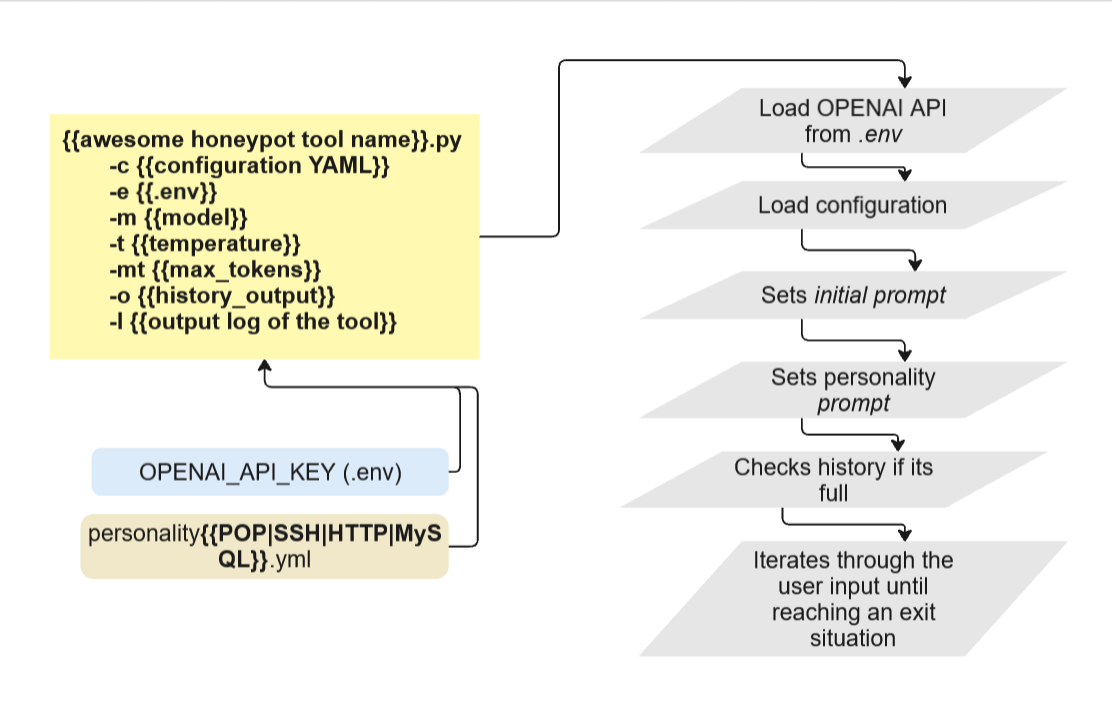}
\caption{The flowchart of behavior and usage of \textit{VelLMes}.}
\label{fig:toolbehavior}
\end{figure}

The flowchart of \textit{VelLMes} components is presented in Figure~\ref{fig:toolbehavior}.

Another reason why the Python script is necessary is to handle special cases and specific commands. For example, when the output should be delivered line by line in specific time intervals. Other cases that are handled by the Python code are conversation exits. Due to the nature of LLMs, the only way to stop a conversation with them is to stop replying to them. Issuing \textit{exit} command in SSH, or equivalent command in other protocols, does not stop the conversation. We solved this by anticipating the correct versions of these commands in the code, waiting for the model to generate a valid response if needed, and then closing the session from the code.

The code for \textit{VelLMes} is provided at the GitHub repository: https://github.com/stratosphereips/VelLMes-AI-Deception-Framework 

\section{Evaluation Experiments}
\label{sec:experiments}

To assess the quality of our proposed solution, we evaluate \textit{VelLMes} in two ways:
\begin{itemize}
    \item Evaluating generative capabilities of LLMs to simulate protocols
    \item Evaluating deceptive capabilities to deceive attackers
\end{itemize}

To evaluate the generative capabilities of LLMs to simulate protocols in \textit{VelLMes} we use \textit{unit tests for LLMs}. To evaluate the deceptive capabilities we conduct an experiment with 89 attackers who need to figure out whether they are interacting with an Ubuntu shell or a \textit{shelLM} honeypot. Furthermore, we deploy 10 \textit{shelLM} instances on the Internet to catch real-life attacks.

These evaluations are described in more detail in the following subsections.

\subsection{Unit tests}
We create, design, and implement a concept of unit tests for LLMs to evaluate the generative performance of cloud-based and local LLMs. This testing helps us to easily, repeatedly, and quickly evaluate the generative abilities of LLMs to simulate various services and not their deceptive capabilities.

The main problem to be solved when designing these unit tests is how to deal with the non-deterministic nature of LLMs. The tests have to evaluate the behavior of the models but still take into consideration that the model response might not be the same on each run. The solution to this issue is to look for the specific parts of the LLM response that are always expected to be in the response.

During the design stage of the unit tests, we create two versions. We do so to check how the conversation history and context can affect the LLM's behavior.

In the first version, all the tests are run one after the other in one unique session, with the whole conversation history of the previous test available to the LLM for each new test. We denote this session type as \textit{Whole}.

In the second version, each test starts a new session with the model and finishes the session at the end, with no history or influence between the tests. This means that each session is separated, with a fresh new context and without the history of the previous test. We denote this session type as \textit{Split}.

The check whether the test was successful or not is a crucial part of our methodology design. The check has to be carefully designed to avoid errors such as false positives and false negatives. Our methodology consists of the following three types of checks:

\begin{itemize}
    \item Existence of the expected substring;
    \item Length of the output;
    \item Consistency between multiple commands' responses.
\end{itemize}

\subsubsection{Setup}

For each of the MySQL, POP3, and HTTP services in \textit{VelLMes} we create 10 ground truth rule tests~\footnote{Detailed unit test descriptions provided at: \href{https://github.com/stratosphereips/VelLMes-AI-Deception-Framework/tree/main/Unit\%20Tests\%20for\%20LLMs}{https://github.com/stratosphereips/VelLMes-AI-Deception-Framework}} which are used to check generative capabilities of LLMs. For the \textit{shelLM} we create 12 tests, due to its complexity, size, and number of areas to evaluate. Each of these test sets can further be extended with new tests.

In the case of \textit{shelLM}, we evaluate how the LLMs behave in scenarios like file creation and manipulation, movement through the file system, responding to invalid commands, and responding to prompt injections. 

For the MySQL honeypot, we evaluate how the LLMs behave in scenarios like navigating through the database, reading, creating, deleting, and updating tables and table fields, and how attempts of prompt injections are handled.

For the POP3 honeypot, we evaluate how the LLMs behave in scenarios like displaying, reading, and deleting messages, and how prompt injections are being handled.

For the HTTP honeypot, we evaluate the ability of LLMs to generate HTML and CSS code, respond to valid and invalid requests, and handle prompt injections properly.

We run unit tests with GPT-3.5 and GPT4 as LLMs behind \textit{VelLMes}. We run both versions of unit tests, \textit{Whole} and \textit{Split}, on all the simulated services and with both GPT-3.5 and GPT-4. For evaluation of \textit{shelLM} we also include the fine-tuned GPT-3.5

\subsection{Expert Evaluation}
\label{sec-expert-evaluation}
To evaluate the deceptive capabilities of simulated Linux shell \textit{shelLM}, we conducted an experiment involving 89 attackers. This experiment was conducted during the university class on computer security, and it involved attackers from all over the world, given that the class is also taught online.

\subsubsection{Setup}

We deployed two shell instances, one running a real Ubuntu system and the other one running \textit{shelLM}. The attackers were assigned randomly, with equal 50/50 probability to one of these shell systems. They were notified in advance that the system they would interact with might be real or a honeypot. We did not mention that the honeypot would be an LLM-simulated shell. 

The experiment was set up on a webpage that contained an embedded terminal through which we connected attackers to the assigned shell. After the interaction was done, the embedded terminal was hidden, so that there would be no possibility of interaction while answering the survey. There was no need for the attackers to do anything outside this webpage. We also provided rules, task explanations, a privacy notice, and the final survey on the same webpage.

The task for the attackers was to exfiltrate a secret crypto wallet key, without being detected. Afterward, they were given a survey in Google Forms to say whether they think they interacted with a real Ubuntu shell or a honeypot, how sure they are in their conclusion, and why they think so.

\subsection{Real-life Attacks}
To evaluate and analyze the behavior of LLM-based honeypots against real attacks on the Internet, we decided to deploy \textit{shelLM} as a real Internet-facing service. Contrary to the previous experiments where human attackers will often follow the guidelines of a normal operation of a shell system, with Internet attacks a system can receive any type of attack that often does not respect the normal operation of the systems. 

\subsubsection{Setup}
We set up 10 identically configured cloud servers running \texttt{Ubuntu 24.10 x64}. The cloud provider chosen was Digital Ocean~\cite{digitalocean}. Each server had 1 CPU, 512 MB Memory, and 10 GB Disk. To reduce the variability between experiments and be able to compare attacks more objectively, all cloud servers were placed in the same data center in Frankfurt. The location was chosen because it is one of the most popular locations for receiving attacks~\cite{VALEROS2025111261}. The SSH port used for the administration of the server was moved from port 22 to port 900 immediately after its creation. \textit{shelLM} was deployed using Docker on port 22/TCP. The experiment was run for 5 days between February 11 and 15, 2025. The LLM logs were regularly copied for analysis. 

In these experiments, \textit{shelLM} used the same prompt, configuration, and LLM model as in the expert evaluation described in Section~\ref{sec-expert-evaluation}. Attackers connecting to port 22/TCP will be prompted for a user and password. Only one accepted combination was valid (\texttt{admin:admin}). Upon successful log-in, attackers interfaced directly with \textit{shelLM}. Contrary to other evaluations, in these experiments, we did not set a goal for the attackers.

\textit{shelLM} accepts only interactive sessions, as these are like doing a manual SSH session to a server. Non-interactive command execution connections are rejected.

\section{Results} 
In this section we present the results of the previously described evaluations.

\subsection{Results of Unit Tests Evaluation}

\begin{figure}
\includegraphics[width=1.0\linewidth]{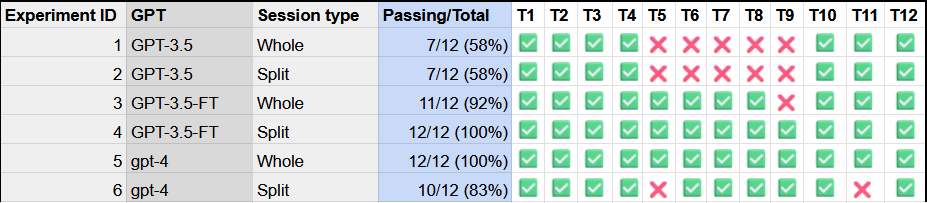}
\caption{Results of Unit Testing the \textit{shelLM} Linux shell Honeypot}
\label{fig:shellmtests}
\end{figure}

\begin{figure}
\includegraphics[width=1.0\linewidth]{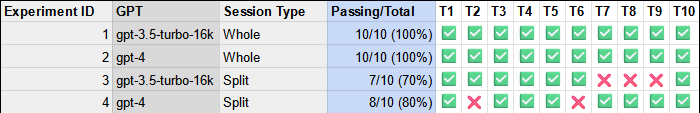}
\caption{Results of Unit Testing the MySQL Honeypot}
\label{fig:msqltests}
\end{figure}

\begin{figure}
\includegraphics[width=1.0\linewidth]{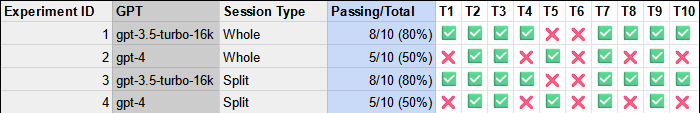}
\caption{Results of Unit Testing the POP3 Honeypot}
\label{fig:pop3tests}
\end{figure}

\begin{figure}
\includegraphics[width=1.0\linewidth]{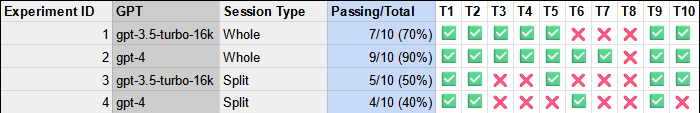}
\caption{Results of Unit Testing the HTTP Honeypot}
\label{fig:httptests}
\end{figure}

The results of unit tests are presented in 4 separate tables, each one for a different service simulated by \textit{VelLMes}. The tables contain information about the LLM used, session type, and passing tests. The results of unit tests for \textit{shelLM}, MySQL, POP3, and HTTP honeypots are shown in Figure~\ref{fig:shellmtests}, Figure~\ref{fig:msqltests}, Figure~\ref{fig:pop3tests}, and Figure~\ref{fig:httptests}, respectively.

In the case of \textit{shelLM}, it can be seen that GPT-3.5 was successful in 7/12 (58\%) tests in both versions of unit tests. The majority of failing tests are the ones that check for prompt injections to verify if the shell is LLM simulated. GPT-4 had a much better performance passing all the tests when they were run as a part of a single session and failing 2 (17\%) tests when they were run each in a separate session. The best performance was by the fine-tuned GPT-3.5 which had only one failing test in the \textit{Whole} session type. In the \textit{Split} session type it passed 100\% of tests.

In the case of MySQL, it can be seen that both GPT-3.5 and GPT-4 were very successful when all the tests were run as part of one session, both passing 10/10 (100\%) tests. When each test was run in a separate session GPT-4 was a bit more successful, with 2 failing tests, than GPT-3.5 with 3 failing tests.

In the case of POP3, GPT-3.5 outperformed GPT-4. In both versions of unit tests, GPT-3.5 had 2 failing tests, while GPT-4 had 5 tests failing.

In the case of HTTP, the most successful LLM was GPT-4 when all the tests were run as a part of a single session, with only 1 failing test. However, when tests were run in separate sessions it had 4/10 (40\%) of successful tests. The GPT-3.5 had 7/10 (70\%) and 5/10 (50\%) tests passing for \textit{Whole} and \textit{Split} unit tests versions, respectively.

From these results we conclude that LLMs can be used to simulate various protocols and services, some more successfully than others. It is also noticeable that the way in which conversation is conducted can greatly affect the successful generation of LLMs' responses.

\subsection{Experts Evaluation Results}

In total 89 attackers took part in the experiment. The real Ubuntu shell was assigned a total of 46 times and \textit{shelLM} honeypot a total of 43 times.

In the case of real Ubuntu 65.2\% of attackers were of the opinion they interacted with a honeypot, and 34.8\% said they interacted with a real system. In the case of \textit{sheLM} results are very similar with 69.8\% of attackers saying they interacted with a honeypot, while 30.2\% thought they interacted with a real system. These results are visible in Table~\ref{tab:conf_matrix}.



\begin{table}[h]
    \centering
    \begin{tabular}{|c|c|c|c|}
        \hline
        \textbf{Actual / Predicted} & \textbf{Honeypot} & \textbf{Real Ubuntu} & \textbf{Total} \\
        \hline
        \textbf{Real Ubuntu} & 30 (65.22\%) & 16 (34.78\%) & 46 (100\%) \\
        \hline
        \textbf{shelLM} & 30 (69.77\%) & 13 (30.23\%) & 43 (100\%) \\
        \hline
    \end{tabular}
    \vspace{10pt}
    \caption{Results of Attacker Evaluation of \textit{shelLM}}
    \label{tab:conf_matrix}
\end{table}

Due to a high percentage of attackers wrongly identifying Ubuntu as a honeypot we find it likely that a certain amount of bias was introduced by mentioning the idea that they might be interacting with a honeypot and by conducting the evaluation during a university class on computer security.

Therefore, we look at the explanations of the reasoning behind the attackers' decisions. 
Some reasons for flagging real Ubuntu as a honeypot were:

\begin{tcolorbox}[halign=center] 
\small
\begin{verbatim}
Many files with juicy names.
The solution was too easy.
Not all bash commands are available.
\end{verbatim}
\end{tcolorbox}

Reasons some attackers gave for flagging \textit{shelLM} as a honeypot were:

\begin{tcolorbox}[halign=center] 
\small
\begin{verbatim}
Some commands worked only sometimes.
I didn't manage to set an alias.
Command ps gave partial answer.
\end{verbatim}
\end{tcolorbox}

Some reasons why attackers thought they interacted with the real system when they were assigned to \textit{shelLM} were:

\begin{tcolorbox}[halign=center] 
\small
\begin{verbatim}
Tried many commands. Seemed normal.
I didn't see anything strange.
Responses of the shell seemed normal.
\end{verbatim}
\end{tcolorbox}

After asking participants whether they were interacting with a honeypot we also asked them to state how sure they are in their choice, on a scale from 1 to 5, where 1 means definitely a real system, and 5 definitely a honeypot. In roughly 30\% of the answers, for both systems, attackers chose option number 3 on the scale, which indicates that they were unsure whether the system was real or not. 

From this experiment, we confirm that LLMs can be used to simulate Linux shells and that they are able to deceive some users interacting with such systems. If we account for the likely bias, we hypothesize that the actual number of attackers claiming they interacted with a real system would be closer to 50\%, with more attackers making such a claim for Ubuntu than for \textit{shelLM}. This is mainly due to some responses of the \textit{shelLM} that were obviously incorrect or hallucinations.

However it is not very easy to verify such a claim. We find it very difficult to design such an experiment without introducing bias, even without mentioning the possibility of a honeypot before the experiment begins. This can be an interesting problem for future research. 

Nonetheless, to the best of our knowledge, this experiment is the largest evaluation of honeypots by real attackers up to now.

\subsection{Real-Life Attacks Results}

Our \textit{shelLM} real-life experiment yielded 2,825 attack sessions. Only 151 were found to be fully interactive sessions valid for analysis. We grouped these 151 sessions according to the attack received, resulting in four attack groups. 

An evaluation of the 151 sessions showed very promising results. The 151 sessions resulted in a total of 276 commands executed. Our analysis showed that \textit{shelLM} answered with a correct output in 98.91\% of these commands. In the remainder of 1.09\% of the commands, the outputs generated by \textit{shelLM} contained obvious mistakes or broken outputs. We found no cases of hallucinations in response to commands. However, we found three hallucinations, each at the end of a session, in which the LLM printed some unnecessary content. 

\subsubsection{Type I Attacks}
Type I attacks were automated attacks generated by a variant of the Mirai botnet. They connected to the SSH and sent a batch of commands that included the download and execution of a binary from the Internet.

We identified 21 Type I attack sessions that resulted in 147 executed commands. The results of the analysis show that in 98.64\% of the commands executed for Type I attacks, \textit{shelLM} produced a correct output.

\subsubsection{Type II Attacks}
Type II attacks were automated attacks against Oracle systems. These sessions consisted of a single automated attempt to change the password and close the session.

We identified 46 Type II attack sessions that resulted in 46 executed commands. The results of the analysis show that in 97.88\% of the commands executed for Type II attacks, \textit{shelLM} produced a correct output.

\subsubsection{Type III Attacks}
Type III attacks were automated attacks that transferred a file to the compromised SSH host and closed the session. We were unable to associate this attack with any known malware to date.

We identified 83 Type III attack sessions that resulted in 83 file transfers. The results of the analysis show that in 100\% of the file transfer commands, \textit{shelLM} produced a correct output.

\subsubsection{Type IV Attacks}
Type IV attacks were manual human attacks. These sessions consisted of a human attacker executing commands. Human attacks are very rare.

We identified 1 Type IV attack session that resulted in 7 executed commands. The attacker appeared to be evaluating the system and executing commands like \textit{ifconfig, ls, crontab, top, history, and ps}. The results of the analysis show that in 6 of 7 commands, 85.71\%, \textit{shelLM} produced a correct output.

\section{Conclusions and future work}
\label{sec:conclusion}
This paper presented \textit{VelLMes} an AI-based deception framework, capable of simulating services such as SSH Linux shell, MySQL, POP3, and HTTP. All of these services can be deployed and used as honeypots. During the design phase, we used several prompt engineering techniques and for the SSH Linux shell we used a fine-tuned LLM. 

The behavior of \textit{VelLMes} was evaluated in three ways. The generative capabilities were evaluated using \textit{unit tests for LLMs}. These tests showed that LLMs are capable of simulating popular protocols and services, with some LLMs passing all the tests. The deception capabilities of the SSH Linux shell were evaluated with 89 human attackers, who had to recognize whether they interacted with a honeypot. Around 30\% of the attackers who interacted with the LLM honeypot mistook it for a real system. Lastly, we evaluated the behavior of the SSH Linux shell in real-life scenarios by deploying it on the Internet and measuring attacks. This evaluation showed that LLMs can handle real-life situations well since the LLM honeypot correctly responded to commands more than 90\% of the time and even managed to engage a human attacker to manually inspect the system.

In future work, we plan to improve the quality of LLM responses to make the deception even more realistic, by experimenting with different and more recent LLMs. The \textit{VelLMs} by design leaves room for more protocols and services to be included and that can be an ongoing improvement to this deception framework.

Furthermore, we plan to design more evaluations both with human attackers and with honeypots being open to real-life attacks. We consider it an interesting and important challenge to find a way to design experiments that would minimize participants' biases and give us even better insight into the usefulness of deception techniques we are using.


\bibliographystyle{IEEEtran}
\bibliography{bibliography}

\end{document}